\documentclass{article}

\usepackage{PRIMEarxiv}

\usepackage[utf8]{inputenc} 
\usepackage[T1]{fontenc}    
\usepackage{hyperref}       
\usepackage{url}            
\usepackage{booktabs}       
\usepackage{amsfonts}       
\usepackage{nicefrac}       
\usepackage{microtype}      
\usepackage{lipsum}
\usepackage{fancyhdr}       
\usepackage{graphicx}       
\usepackage{cite}
\usepackage{amsmath,amssymb}
\graphicspath{{media/}}     

\title{Automating SBOM Generation with Zero-Shot Semantic Similarity
}

\author{
  Devin Pereira, Christopher Molloy, Sudipta Acharya, Steven H.H. Ding \\
  School of Computing \\
  Queen's University \\
  Kingston\\
  \texttt{\{devin.pereira, chris.molloy, s.acharya, steven.ding\}@queensu.ca} \\
  }

\begin{document}
\maketitle

\begin{abstract}
It is becoming increasingly important in the software industry, especially with the growing complexity of software ecosystems and the emphasis on security and compliance for manufacturers to inventory software used on their systems. A Software-Bill-of-Materials (SBOM) is a comprehensive inventory detailing a software application's components and dependencies. Current approaches rely on case-based reasoning to inconsistently identify the software components embedded in binary files.  We propose a different route, an automated method for generating SBOMs to prevent disastrous supply-chain attacks. Remaining on the topic of static code analysis, we interpret this problem as a semantic similarity task wherein a transformer model can be trained to relate a product name to corresponding version strings. Our test results are compelling, demonstrating the model's strong performance in the zero-shot classification task, further demonstrating the potential for use in a real-world cybersecurity context. 
\end{abstract}


\section{Introduction}
Throughout the past decade, major security breaches have been caused by malicious code embedded in firmware, a type of software that directly interacts with hardware. Attacks on firmware can be particularly revealing and destructive, as firmware has high privileges over other aspects of computer systems. Recently, the devastating effects of exploiting firmware vulnerabilities were evidenced by the SolarWinds supply chain attack \cite{lazarovitz2021deconstructing}, an attack that compromised thousands of organizations around the world, including the United States government.

A supply chain attack constitutes a security breach where a malicious entity exploits vulnerabilities within third-party software to gain unauthorized access to the targeted system rather than attempting a direct breach of the target system itself. Supply chains associated with contemporary software systems are evolving into more intricate structures, contributing to increased opacity and making them susceptible to such attacks. It follows that it has become extremely difficult to identify which components are present in a particular piece of software \cite{muiri2019framing}. The problem finds its source in the software engineering practice of code reuse. Since this custom is unlikely to be abandoned, an alternative approach to prevent supply chain attacks is required.

In 2018, the National Telecommunications and Information Administration of the United States Department of Commerce (NTIA) formed a working group that proposed the Software-Bill-of-Materials (SBOM).
Concisely, an SBOM provides details about all the components present in a given piece of software. SBOMs detail information such as the software supplier, the component names and versions present, as well as the relationships between these components.

However, it is important to note that SBOMs do not guarantee security throughout the supply chain. An SBOM is not applicable against zero-day exploits, nor can it be used to flag a security flaw before it is reported to a vulnerability database \cite{phillips2023software}. Nevertheless, the simplest approach to derive an SBOM is to use the relevant package manager to list a software’s dependencies. Though simple and highly effective, it requires knowledge of the package manager used for installation. Evidently, the technique does not work for software installed without a package manager. 

The alternative procedure of deriving an SBOM from a software’s source code is equally reliable. However, the main limitation of this approach is that source code is not guaranteed to be available. In general, it is near impossible to locate source code for a specific binary, particularly in the case of firmware, which often is preinstalled on systems \cite{phillips2023software}. There is a salient need for a tool capable of producing SBOMs using only compiled binaries. Such a tool could guarantee the integrity of vendor-generated SBOMs and protect against tampering by adversaries, i.e., a man-in-the-middle attack in which a software is maliciously modified, but its SBOM deceitfully remains the same. 

Henceforth, we refer to SBOM generation tools as utilities for deriving SBOMs from binaries. Tools like these exist but have the drawback of relying on case-based reasoning. In fact, many of them require predetermined unique patterns for matching strings to a specific set of software products. 

To remedy this issue at scale, we elect to implement a machine-learning solution for matching version strings to software products. We demonstrate that this problem can be solved using modern natural language processing techniques. Our transformer model is tasked with determining the semantic similarity between a software product name and the version strings embedded in software binaries. The most similar software product to the version string is then selected as the network's prediction. We intend that these predictions be used as input for the automated generation of SBOMs and to query vulnerability databases if there are CVEs corresponding to that particular product. We show our model’s ability for zero-shot inference by testing its detection accuracy on multiple software versions not encountered in training. Our contributions can be summarized as follows:
\begin{itemize}
  \item We propose the first transformer model optimized for the SBOM generation task. As opposed to a naive pattern-matching operation, identifying software modules is formulated as a semantic-similarity-based classification task. 
  \item By way of zero-shot inference experiments, we demonstrate that our model, despite being trained on a few classes, can be applied successfully with only a small degradation in performance when compared to a state-of-the-art model trained on all available classes. The outcome underscores our method as being effective in practical cybersecurity contexts.
\end{itemize}
The paper's structure is as follows. The \emph{Related Works} section highlights a series of studies from which we draw inspiration. The \emph{Similarity Learning for Version Detection} section contains the preparation of our dataset, a description of the data employed, and a illustration of our model. In \emph{Experimental Setup and Results} section, we explain our experiments and describe our interpretation of the results. Finally, the paper concludes and provides potential scopes for future works in \emph{Conclusion and Future works} section.

\section{Related Works}
In this section, we review the current tools for generating SBOMs. We appraise studies wherein transformer models have been effectively employed for pattern detection. Finally, we present approaches to classification tasks which involve zero-shot learning.

\subsection{SBOM Generation Tools}
Intel's open source CVE-bin-tool is a widely used software analysis program which employs a binary scanner to determine which software versions are embedded in a binary file and subsequently maps these versions to known vulnerabilities in the National Vulnerability Database \cite{booth2013national}. Doan and Jung employ CVE-bin-tool to develop DAVS–A Dockerfile analysis-based vulnerability scanning framework \cite{doan2022davs}. 

Their approach aims to create a methodology for vulnerability detection that is not restricted by the ability of package managers to extract information. After determining potentially vulnerable files through preliminary dockerfile analysis, they make use of CVE-bin-tool to detect software versions, linking them to vulnerabilities. Reinhold \textit{et al}. compare the results produced by different versions of CVE-bin-tool on 660 binary files \cite{reinhold2023new}. They observe that only $1$ out of the $660$ binaries (excluding null results) produces an output consistent across the different versions the tool. This outcome raises additional doubts about the reliability of the case-based approach and underscores the necessity for a new paradigm.

\begin{figure*}[t]
\centering
\includegraphics[width=0.8\textwidth]{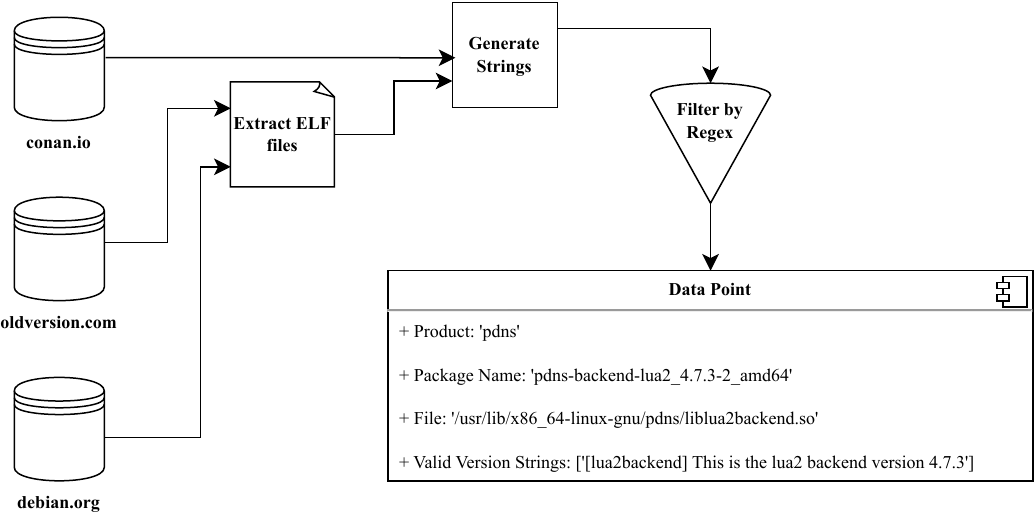} 
\caption{Binaries obtained from different sources are passed through extraction scripts. Next, strings are generated from the resultant ELF files. Finally, these strings are funneled through a regular-expression based filtering script. The final data points can be observed as a tuple with four different fields.}
\label{figure1}
\end{figure*}
\subsection{Transformer Models}
Very little research addresses the problem of determining the similarity between pseudo-English words. Such is the nature of the task in comparing a collection of version strings derived from binary files to software names. However, the current state-of-the-art class of models has been used extensively for classification tasks, which we consider to be a related problem type to semantic similarity. Shaheen \textit{et al.} demonstrate the efficacy of using transformer models in text classification \cite{shaheen2020large}. To achieve state-of-the-art results in Large Multi-Label Text Classification (LMTC), they compare different models alongside various strategies such as generative pretraining, gradual unfreezing, and discriminative learning rates. Across the different datasets, BERT (Bidirectional Encoder Representations from Transformers) variants were observed to yield the best results. Interestingly, the researchers note that multilingual BERT trained only in English data yields results on par with the others when presented with texts from other languages. This indicates BERT’s zero-shot learning capabilities.

BERT is the current state-of-the-art embedding model for natural language introduced by Devlin \textit{et al}. \cite{devlin2018bert}. Their novel bidirectional attention model allows for learning information from a text both left to right and right to left. In order to develop a deep bidirectional representation, the researchers mask some percentage of the input tokens at random during training and then predict the masked tokens. This procedure is known as masked language modeling (MLM). Prior left-to-right-only approaches restricted models to unidirectional learning. BERT has shown to be seminal as it can be used as a foundation for other models. Fine-tuning can be achieved with as little as one additional output layer to create state-of-the-art models for specialized natural language tasks.

In the year following the release of BERT, Nils Reimers and Iryna Gurevych proposed a method to increase the efficiency of BERT in a sentence similarity task. \cite{reimers2019sentence}. The motivation behind their improvement relates to a shortcoming of BERT: its cross-encoder setup is unsuitable for a variety of pair regression tasks due to the excessive number of possible output combinations it allows. Instead, the researchers develop a Siamese network architecture which enables the derivation of fixed-sized vectors from input sentences. These fixed-size vectors can be viewed as semantically meaningful sentence embeddings. They assert that, in a task to determine the similarity between $10,000$ sentences, the necessary computation time is reduced from $65$ hours with BERT, to around $5$ seconds with S-BERT. In brief, their improved model drastically reduces the computational overhead, thereby improving the time taken to determine the most semantically similar pair of sentences. Simultaneously, their method almost perfectly maintains the near stellar accuracy of BERT.

\subsection{Zero-shot Learning}
The formalization of zero-shot learning is attributed to Palatucci \textit{et al}. These researchers introduce a so-called semantic output code classifier, which infers a knowledge base of semantic properties from samples in a training set. They demonstrate that their classifier can accurately predict outputs from novel classes, emphasizing the relevance of the task when outputs have the potential to take on a large number of different values, as well as when the cost of obtaining labeled data is high. The model used in their experiments is made to predict words using functional magnetic resonance images from human participants. They justify the necessity of their novel procedure by stating that determining neural training images for every English word in existence is intractable. Over two semantic knowledge bases, human218 and corpus5000, their method obtains mean accuracies of $80.9\%$ and $69.7\%$, respectively. These results were shown to be powerful, despite originating from the brain waves of only nine participants. From its inception, the concept of zero-shot learning was linked to classification tasks with a semantic character. \cite{palatucci2009zero}

Ziming Zhang and Venkatesh Saligrama describe the zero-shot learning problem from an alternative perspective. They view target data instances as arising from observed instances, attempting to express source and target data as a mixture of observed class proportions. Their idea is, if the mixture proportion from the target domain is similar to that from the source domain, the target and source necessarily originate from the same class. Consequently, they formulate the problem as learning source and target domain embedding functions using observed class data. These functions are designed to map arbitrary source and target domain data into mixture proportions of observed classes.
Their method involving semantic similarity embedding functions improves the state-of-the-art accuracy CIFAR-10 to $88.3\%$ amongst other benchmarks. \cite{zhang2015zero}

\section{Similarity Learning for Version Detection}

\subsection{Data Preparation}

We require a dataset of varying software products and version strings for training our network. We collect product and version information from Portable Executable (PE) and Executable and Linkable Format (ELF) file type binaries. The PE format used by Windows specifies the structure of executable files (.exe) and Dynamic Link Libraries (DLLs), which are linkable software packages. Though there are many other file types which fall under the PE umbrella, we collect only these two in our dataset. Note that these are designed for many architectures e.g. x86, arm, amd64, etc. Analogous to PE but conceived for Unix systems, the ELF format is commonly used for executables, object code, shared libraries, and more.

We obtain our files from two sources: conan.io and debian.org. In the case of conan.io, we query the \emph{conan-center} to download files from over $1500$ different libraries. We select only the PE and ELF files from the downloaded directory. This list of files is compared to the reference file paths in the cache to avoid duplication, renamed, and finally placed in the dataset. The procedure is nearly identical for the debian packages.

Once the EFL files originating from all sources are gathered, they are transformed into strings, simply using the Unix \emph{strings} command. Within these strings, we consider only the strings that match a general regular expression or version string pattern. From the metadata in the repository, we obtain a package name that contains the version of the software. This version number is used as a ground truth to label each version string. The product is simply the name from the conan or debian package catalogue.

\subsection{Data Description}

Though SBOMs contain many other fields, we deem most of these irrelevant to our research. The SPDX format, the standard structure for SBOMs, was originally conceived to manage product licensing information \cite{Administration_2021}. Hence, many of these fields are not pertinent to vulnerability analysis which we consider to be the primary purpose of SBOMs in our study. For example, relationships between components are not recorded, our method only identifies the components which are present in the piece of software. 

There are other fields which our approach does not directly cover. Manufacturers can be easily extrapolated from the product name. As mentioned above, version numbers, which are firmly required for SBOM generation and CVE identification, are used to label the version strings and thus are not subsequently determined by the machine learning solution. 

In summary, each data point in our final processed dataset, illustrated in Figure \ref{figure1}, contains a product, a package name and a list of valid version strings. Finally, having already correlated each valid version string to the product, we decorrelate those version strings with the other unrelated products to optimize future classification.

\subsection{Model}

\begin{figure}[t]
\centering
\includegraphics[width=0.4\columnwidth]{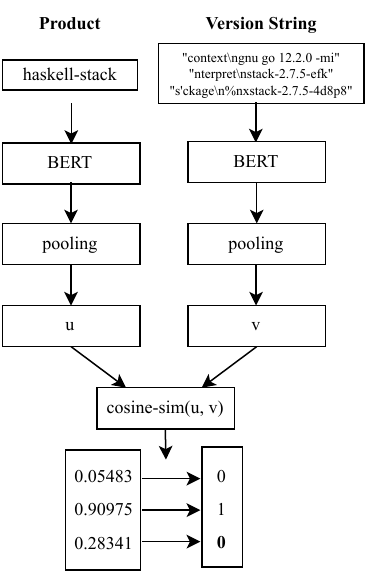} 
\caption{The product and one version string from the list is taken as a data point which serves as input to the S-BERT model. S-BERT produces embeddings which are pooled into two vectors, u and v. The cosine similarity is calculated with these two vectors and contrasted with the predefined correlation label to obtain the final classification}
\label{figure2}
\end{figure}

In BERT and S-BERT, inputs are parsed according to the WordPiece tokenization algorithm. We will now describe the WordPiece tokenization algorithm \cite{wu2016google}. As an initial step, a WordPiece model must be generated through the training procedure. It can be explained as follows: the model is initialized with wordpieces as word units or simple characters. Given a training corpus with stopwords filtered, the model does a pairwise combination of wordpieces to increase the likelihood of them being present in the training corpus. This process is continued until the minimum number of wordpieces for the corpus is constructed. Alternatively, training may be made to stop once a predefined number of wordpieces is attained, in which case the remainder will be out-of-vocabulary. 

Prior to training, word separators are added such that the original sequence of words can be recovered from the wordpiece sequence produced by the model. BERT uses two special tokens in its version of the WordPiece algorithm. The special classification token [CLS], whose final hidden state is used as the aggregate sequence representation in the classification task, and the [SEP], which delineates the two inputs. As well, many rare characters are not represented to minimize inefficiencies. Instead, these are replaced by an unknown placeholder character ([UNK]). An additional constraint is that input sentences longer than $256$ word pieces are truncated. Notably, this approach remains unaffected by word or sentence semantics. 

In the context of our model, the WordPiece tokenizer can be described as a mapping from $\mathbb{Z}^{U*l} \rightarrow  \mathbb{Z}^{W*l}$ where $U$ is the number of unicode representations, $l = 256$ is the maximum input sequence length and $W = 30,000$ is the number of wordpieces. These vectorized wordpieces are then combined with positional embeddings and binary segment embeddings to indicate provenance from the first or second sentence. The procedure can be observed as $\mathbb{Z}^{W*l} \rightarrow  \mathbb{R}^{H*l}$ where $H = 768$ is the hidden size or the size of the embeddings for each wordpiece. Subsequently, these BERT embeddings are transformed according to the mean-pooling operation $\mathbb{R}^{H*l} \rightarrow  \mathbb{R}^{e}$ where $e = 384$ is a dense vector space representing the fixed-size of the S-BERT embedding vector which captures the entire input sequence. These embeddings are used as inputs to an objective function $S(x)$ from $\mathbb{R}^{e} \rightarrow  0 < \mathbb{R} < 1$. The resulting real number is the probability that a given product name corresponds to the queried version string. 

\begin{table*}
\centering
\caption{All $20$ classes present in the training set. The training set contains $4,000$ samples from each.}
\begin{tabular}{|l|l|l|l|l|}
agda & ghc & gcc-xtensa-lx106 & fricas & gcc-13 \\
grass & haskell-opengl & haxml & gcc-12-cross-mipsen & grub2 \\
pandoc & haskell-gtk & haskell-graphviz & gcc-riscv64-unknown-elf & haskell-gtk3 \\
darcs & metaeuk & uuagc & gcc-10-cross & propellor \\
\end{tabular}
\label{table1}
\end{table*}

To compute the sentence similarity score from two inputs, we employ an S-BERT implementation from the Sentence Transformers package, specifically the \emph{all-MiniLM-L6-v2} model. It is structured with three components as highlighted above: 1) the BERT model, 2) the pooling layer, 3) the similarity function. Our model employs BERTBASE, which has $12$ standard multi-layer bidirectional Transformer layers, $12$ self-attention heads, and a hidden size of $768$ comprising 110M total parameters. With up to $256$ outputs of hidden size emanating from this network, the mean-pooling layer condenses the information into a fixed-size embedding vector. We calculate similarity scores using these embeddings as input. Finally, the cross-entropy loss is computed on the cosine-similarity score and the predicted label.

\begin{equation}
cosine\_sim(u,v) =  \frac{u \cdot v}{\|u\| \cdot \|v\|}
\end{equation}

In a deployed setting, we expect this tool to be used by a stakeholder who is likely to be a security analyst or a systems administrator. In the first phase, the user must gather as many software binaries as is available to their organization. They must then process the files in their repository following our pipeline to obtain products and a corresponding list of version strings. Importantly, product names must be made consistent for optimal performance. Following our methodology, the model can be trained and tested. At this point, it is ready for practical use. One simple use case involves an organization wanting to install new software in its system. This tool can be relied upon to decide whether the software can be whitelisted. If, after analysis, the model determines that none of the version strings in the software file correspond to a product linked with CVEs, then the software is safe to install. Optionally, the file can be added to the dataset for future training.

\section{Experimental Setup and Results}

We obtain the raw data, a collection of over 100GB of binaries, from various online repositories. From the raw data, we curated a dataset of $400,000$ version pattern samples sourced from a wide range of software binaries.

\subsection{Experiments}
To validate the proposed methodology, we conducted three experiments. The first experiment serves to verify the usefulness of semantic similarity for making predictions about our data. Also, it is designed to guide our selection of the similarity metric to use for later experiments. The second experiment compares the model's efficacy across two different data sampling techniques. It contrasts a perfect simulation run in the lab with a demonstration of how the tool might be employed in practice. We use a training set with significantly less sample diversity to account for the low number of malware samples available to the average user in comparison to the abundance of software ecosystems. Through the third experiment, we determine the optimal training length for our model when used in a practical setting.

We limit the size of the experimental dataset to $100,000$ samples in all experiments. We deem this number to promote both sufficient sample diversity and training efficiency while maintaining the quality of the results, which can be achieved on the entire dataset. Following a similar line of thinking, we consistently split the data into training ($80\%$) and testing ($20\%$) sets.

\begin{table}[t]
\centering
\caption{An extract of the classes in the testing set.}
\begin{tabular}{@{}lllll@{}}
\toprule
Class Name & N Samples & Density \\ \midrule
systemd & 86 & 0.430\% \\
libreoffice & 72 & 0.360\% \\
espresso & 39 & 0.195\% \\ 
medusa & 13 & 0.065\% \\
js-of-ocaml & 3 & 0.015\% \\
scrypt & 1 & 0.005\% \\
xmppc & 1 & 0.005\% \\
\bottomrule
\end{tabular}
\label{table2}
\end{table}

\begin{table}[t]
\centering
\caption{The performance of the model when comparing two different similarity metrics in the zero-shot experiment.}
\begin{tabular}{@{}lllll@{}}
\toprule
Similarity Metric & Accuracy & Precision & Recall  &  \\ \midrule
Cosine            & 0.9290   & 0.9447    & 0.9126  &  \\
Dot Product       & 0.9267   & 0.9452    & 0.90687 &  \\
\bottomrule
\end{tabular}
\label{table3}
\end{table}

\begin{table*}[t]
\centering
\caption{The results compared by inference type. Fully trained in the case of experiment one where the model is trained on all classes. In the zero-shot experiment, our model is trained on only $20$ classes.}
\begin{tabular}{@{}llllll@{}}
\toprule
Inference Type & Accuracy & AUC & Precision & Recall & F-1 Score \\ \midrule
Fully-Trained & 0.9290 & 0.9810 & 0.9447 & 0.9126 & 0.9284 \\
Zero-Shot & 0.8518 & 0.9016 & 0.9512 & 0.7410 & 0.8330 \\ \bottomrule
\end{tabular}
\label{table4}
\end{table*}

\begin{table*}[t]
\centering
\caption{The zero-shot results in different training runs over increasing epochs demonstrate overfitting.}
\begin{tabular}{@{}llllll@{}}
\toprule
Epoch & Accuracy & AUC & Precision & Recall & F-1 Score \\ \midrule
1 & 0.8518 & 0.9016 & 0.9512 & 0.7410 & 0.8330 \\
2 & 0.8225 & 0.8783 & 0.9967 & 0.6463 & 0.7841 \\
5 & 0.7383 & 0.7733 & 0.9777 & 0.4866 & 0.6498 \\
10 & 0.5579 & 0.6267 & 0.6901 & 0.2068 & 0.3182 \\ \bottomrule
\end{tabular}
\label{table5}
\end{table*}

In the first experiment, we randomly select and assign samples from all classes to the two subsets, with the possibility of both sets containing samples from the same class. Then as described above, our S-BERT variant processes this data with a batch size of $512$, moulding embeddings as part of the training process. These embeddings are combined using either cosine or dot product similarity, producing a semantic similarity score, forming the basis for our comparison metrics displayed in Table~\ref{table3}.

The second experiment makes use of two different data selection procedures. In the fully-trained approach, the data is split with samples randomly selected from all classes in the same way as the first experiment. The zero-shot method begins by grouping all samples by class or library. These classes are sorted by the number of samples in descending order to facilitate the filling of the training set.

To match the predetermined size of the training set, $4,000$ samples from each of the $20$ largest classes are drawn and exhibited in Table~\ref{table1}. To the testing set, we randomly allocate samples from the remaining classes in the initial dataset. A subset of the constituents to the testing set can be observed in Table~\ref{table2}. Again the model produces similarity scores, but this time only using dot product similarity. We amalgamate these scores to compute a variety of metrics qualifying the effectiveness of our zero-shot method. These metrics can be seen in Table~\ref{table4}.

In the third experiment, following the same zero-shot procedure as in the second experiment, we conducted four zero-shot trials over an increasing number of epochs. The results can be observed in Table~\ref{table5}.

The first experiment demonstrates that results are largely insensitive to the choice of similarity metric. However, as expected, the cosine similarity metric used to train the model performs best in testing. It outperforms dot product similarity in both accuracy and, notably, recall. Due to minor variations in data selection, it could be observed in a similar experiment that the accuracy of dot product similarity slightly outperforms that of cosine similarity. In this context of cybersecurity, where prevention is paramount, we would still advocate choosing cosine similarity if it captures more positive samples, thereby giving rise to a higher recall score.

\begin{equation}
\sum_{i=0}^{n = 384}u_i\cdot v_i= u_0\cdot v_0 + u_1\cdot v_1 + ... + u_{n-1}\cdot v_{n-1}
\end{equation}
\begin{equation}
u \cdot v = |u|\cdot|v|\cdot cos(\alpha) 
\end{equation}

Results from the second experiment indicate that the fully-trained model performs well, as is our expectation when resented with training and testing data sampled from the same classes. But importantly, we find the model's performance in the zero-shot experiment to be remarkable. Despite training on only $20$ classes in the zero-shot task, the model's accuracy differs by a little over $7\%$ in comparison to the first experiment. Evidently, the recall score suffers from the poverty in sample diversity, leading to the model's inability to recognize as many positive samples. Upon closer inspection, there seems to be a particular maximum number of samples per class or a specific number of classes that is optimal for the zero-shot performance. We determined a local maximum for performance at $4,000$ samples and $20$ classes. 

The outcome of the third experiment concerning the number of epochs used for training is quite simple. The original S-BERT model, itself a fine-tuned version of BERT, was trained over a single epoch. Hence, it is logical that the best results would be obtained by fine-tuning S-BERT over one epoch. In this instance, anything more causes a degradation in performance due to overfitting.

We demonstrate through our experiments that our approach is a solution to the problem of inconsistent results highlighted by Reinhold \textit{et al}. Provided that data is processed with care, any number of different version strings when correctly classified, will consistently match to the same product. Furthermore, existing tools, such as cve-bin-tool and EMBER, only support around $300$ packages, whereas we were able to apply the model to the whole debian repository, which contains over $40,000$ packages. From a practical perspective, this paper describes a method for determining software products. Though it uses version numbers, it does not directly address the problem of identifying them within these products. Nonetheless, version numbers are an important aspect of SBOMs. In fact, they are critical for identifying vulnerable components, as vulnerability databases require version numbers to accurately relate a software component to a vulnerability. Be that as it may, we deem the problem of version number identification to be a trivial operation involving regular expressions. Therefore we have not considered that problem in this paper. 

Reflecting further upon the second experiment, some of the training classes may appear to have version strings with similar structures, for example, haskell-gtk and haskell-gtk3. The relationship between the semantic similarity of version strings from different products in the training set, and zero-shot test performance should be a primary focus of future investigation. At first glance, the solution to the problem might be formulated as a fully connected graph, where the vertices are a subset of the products available, curated for the training set such that the edges are the similarity scores producing a minimum sum for the dataset. Such analysis might be supplemented with a closer look at test results which naturally reveal differences in performance by class. Observing the structure of version strings in the classes which yield low recall scores may also guide the selection of classes in forming an optimal training set.

\section{Conclusion and Future Works}
The purpose of this study is to address the critical issue of supply-chain attacks, whose notoriety was cemented by the Solarwinds supply-chain attack. Recognizing the critical nature of SBOMs in ensuring supply chain security, we brought forward a novel solution that leverages machine learning, specifically the transformer model S-BERT, to automate the generation of SBOMs from compiled binaries. Existing SBOM generation tools, which are generally reliant on case-based reasoning, have demonstrated limitations in dealing with the diverse and dynamic nature of modern software productions.

In an effort to address these challenges, we introduced a method which employs S-BERT to match different version strings present in binaries to the appropriate software product. Three experiments were conducted. The first guided our choice of similarity metric. The second contrasted an instance of our model trained on all classes to one trained on a reduced set to simulate real-world scenarios with zero-shot learning. The third addressed the impact of the number of epochs on the model's performance, emphasizing the risk of overfitting. 

Results showcased the model's impressive zero-shot inference capabilities as it achieved strong results. Though, our findings highlight the importance of sample diversity, raising the question of how class selection should occur for forming training sets. As an open problem for subsequent research, we leave optimizing class selection for training set composition. For a more comprehensive approach to supply chain security, the identification of version numbers within software products is a simple yet important endeavor to undertake in the future. We note that this component is the only other data point required to generate a reduced SBOM sufficient for CVE detection. In conclusion, our research has the potential to protect large scale software supply chains using very little data, thereby making it an extremely valuable tool for protecting organizations and individuals from malicious actors.

\bibliographystyle{unsrt}  
\bibliography{main}

\end{document}